\documentclass[aps,preprintnumbers,nofootinbib,floatfix,11pt]{revtex4}

\usepackage{amsfonts,amsmath,amssymb}
\usepackage{graphicx,epsfig}

\def	\be	{\begin{equation}}
\def	\ee	{\end{equation}}
\def	\bqt	{\begin{quote}}
\def	\eqt	{\end{quote}}

\def	\mn 	{\mu \nu}

\def	\lf	{\left (}
\def	\rt	{\right )}

\def	\comma	{\quad , \quad}

\begin{document}

\title{Thermodynamic Origin of the Null Energy Condition}

\author{Maulik Parikh and Andrew Svesko}

\affiliation{Department of Physics and Beyond: Center for Fundamental Concepts in Science\\
Arizona State University, Tempe, Arizona 85287, USA} 

\begin{abstract}
\begin{center}
{\bf Abstract}
\end{center}
\noindent
We derive the classical null energy condition, understood as a constraint on the Ricci tensor, from the second law of thermodynamics applied locally to Bekenstein-Hawking entropy associated with patches of null congruences. The derivation provides evidence that the null energy condition, which has usually been regarded as a condition on matter, is fundamentally a property of gravity.
\end{abstract}

\thispagestyle{empty}

\maketitle

\noindent
\emph{Introduction.} \textendash{} The null energy condition (NEC) plays a critical role in classical general relativity. It is used in proving a host of gravitational theorems, from the area theorem that states that classical black holes cannot shrink \cite{bardeen}, to singularity theorems that guarantee the existence of the Big Bang \cite{hawkingpenrose}. The NEC is also invoked in excluding bouncing cosmologies and exotic spacetimes containing traversable wormholes and time machines, which might otherwise be exact solutions of Einstein's equations \cite{molinaparis,nobounce,Hawking:1992,Farhi:1987,Morris:1988}. And in asymptotically AdS spaces, the validity of the NEC is equivalent to a c-theorem in the holographic dual theory \cite{warner}. The NEC is usually expressed as the condition
\be
T_{\mn} v^\mu v^\nu \geq 0 \; ,	\label{TNEC}
\ee
where $v^\mu$ is any light-like vector. Here $T_{\mn}$ is the energy-momentum tensor of matter, suggesting that the NEC should be a property of matter. However, our best framework for describing matter -- quantum field theory -- does not appear to have a consistency requirement of the form of (\ref{TNEC}), even as a classical limit. Moreover, several explicit examples of effective theories that violate (\ref{TNEC}) but that are nevertheless not in manifest conflict with the principles of quantum field theory are now known. Thus the origin of a vitally important aspect of general relativity has been mysterious. With no apparent fundamental principle from which the NEC flows, the validity of the NEC has been called into question \cite{Barcelo:2002,Rubakov:2014}.

Motivated by this failure to derive the NEC in some classical limit of quantum field theory, it has been proposed that the NEC should be regarded as a property not purely of matter but of a combined theory of matter and gravity \cite{NECderivation}. In such a theory, Einstein's equations imply that the NEC can be reformulated in a quite different, though equivalent, form as
\be
R_{\mn} v^\mu v^\nu \geq 0 \; ,	\label{RNEC}
\ee
where $R_{\mn}$ is the Ricci tensor. This is now a constraint on spacetime geometry, rather than on energy densities; indeed, it is this geometric form of the null energy condition, known as the Ricci or null convergence condition, that is ultimately invoked in gravitational theorems. Recently it has been shown that precisely this condition can be derived from string theory \cite{NECderivation}, which of course is a theory of both matter and gravity. For a closed bosonic string propagating in an arbitrary graviton-dilaton background, the Virasoro constraints of the effective action lead precisely to (\ref{RNEC}) in Einstein frame, including even the contractions with null vectors. This is a very satisfying derivation of the null energy condition for a number of reasons: It is another example of the beautiful interplay between worldsheet and spacetime, the Virasoro constraints are none other than Einstein's equations in two dimensions, and there is a physical principle -- worldsheet diffeomorphism invariance -- that is associated with the null energy condition, which until now had been an ad hoc condition lacking a clear origin.

Here we shall derive the NEC in an entirely different way. Our premise is that gravity emerges from the coarse-graining of some underlying microscopic theory. This is perhaps a more speculative starting point than string theory, but the derivation has its appeal because it relies on a universal theory, namely thermodynamics. A relation between thermodynamics and the null energy condition is already present in black hole physics. Recall that the NEC is used in deriving the second law of thermodynamics for black holes \cite{bardeen}. The logic runs as follows:
\be
T_{\mn} v^\mu v^\nu \geq 0 \Rightarrow R_{\mn} v^\mu v^\nu \geq 0 \Rightarrow \dot{
\theta} \leq 0 \Rightarrow \theta \geq 0 \Rightarrow \dot{A} \geq 0 \Rightarrow \dot{S} \geq 0 \; .
\ee
Here $\theta$ is the expansion of a pencil of null generators of a black hole event horizon and the dot stands for a derivative with respect to an affine parameter, which can be thought of as time. The first arrow follows from Einstein's equations, the second from the Raychaudhuri equation, the third from avoidance of horizon caustics, the fourth from the definition of $\theta$, and the last from the definition of Bekenstein-Hawking entropy. Ideally, we would like to able to reverse all these arrows so that the NEC flows from the second law of thermodynamics, rather than the other way around \cite{Chatterjee:2012}. However, although the first and last arrows can readily be reversed, provided we assume Einstein gravity and the validity of the gravitational equations, the remaining arrows do not appear reversible. In particular, a serious problem with reversing the arrows is that the second law is a global statement, whereas the NEC is a local condition. 

However, in an ingenious paper \cite{einsteineqnofstate}, Jacobson was able to obtain Einstein's equations, which are also local, from essentially the first law of black hole thermodynamics. The key idea was to assume that, in keeping with the universality of horizon entropy, the first law could be applied to local Rindler horizons. Thus a global law was ``gauged," which was a pre-requisite for obtaining the local gravitational equations of motion. In the same vein, we will show that the null energy condition too, in the form of the Ricci or null convergence condition, (\ref{RNEC}), comes out of thermodynamics applied to a local holographic screen. In a nutshell, just as Jacobson regarded the {\em first} law as an input and obtained Einstein's equations as an output (reversing the laws of black hole mechanics, as it were), we shall regard the {\em second} law as an input and obtain the null energy condition as an output.

Note that we will consider only the classical null energy condition. Much effort in the literature \cite{Wall:2009,inviolable,kontou,QNEC,faulkneretal,hartmanetal} has been directed at proving a quantum null energy condition, $\langle T_{\mu \nu} \rangle k^\mu k^\nu \geq 0$, or generalizing the concept to some kind of averaged null energy condition. Indeed, the standard null energy condition is known to be violated even by Casimir energy. So why focus on the classical NEC? First, the properties of the classical stress tensor are of independent interest. Typically, whenever exotic matter is proposed  in the literature e.g. phantom fields, galileons, ghost condensates, etc., the gravitational consequences are worked out by coupling Einstein gravity to the classical stress tensor of such matter. So it is important to prove the generic properties of this tensor. Second, in attempts to prove the quantum null energy condition, the validity of the classical NEC is often assumed -- yet this needs to be proven. Third, it is not obvious that the expectation value of the quantum stress tensor, as computed, has any gravitational consequences. A quantum null energy condition $\langle T_{\mu \nu} \rangle k^\mu k^\nu \geq 0$ would certainly be meaningful if there were a semi-classical Einstein equation of the form $G_{\mu \nu} = 8 \pi G \langle T_{\mu \nu} \rangle$. However, such an equation is not known to have any rigorous derivation. By contrast, whatever be the ultimate theory of quantum matter coupled to quantum gravity, it surely admits a well-defined $\hbar = 0$ limit of classical gravity coupled to classical matter, which is the situation considered here.

\emph{From the Second Law to the NEC.} \textendash{} 
Before entering into the details, let us summarize the logic of the derivation. First we will quote a statistical-mechanical result about the non-positivity of the second time-derivative of entropy. This is a very general result which holds for virtually all near-equilibrium thermodynamic systems. Next we will propose a prescription for associating thermodynamic systems to patches of null congruences in spacetime. Then will then show that, in the vicinity of any point in spacetime, null congruences corresponding to near-equilibrium thermodynamic systems can always be found. By the quoted result, these then necessarily have non-positive second time-derivative of entropy. Finally, substituting this into the Raychaudhuri equation will imply the Ricci convergence condition, (\ref{RNEC}), which is the geometric form of the null energy condition.

Consider then a finite thermodynamic system and let $S_{\rm max}$ be its maximum coarse-grained entropy. For systems already at equilibrium, $S = S_{\rm max}$, and $\dot{S}, \ddot{S} = 0$. For systems approaching equilibrium, $S < S_{\rm max}$ and the second law says that $\dot{S} \geq 0$. Now, since the entropy tends to a finite maximum value as it approaches thermal equilibrium, and since $\dot{S}\geq0$, it seems intuitively reasonable that the first time derivative of entropy will be a decreasing function of time: $\ddot{S}\leq0$. This inequality, which will be crucial below, indeed holds for a great many systems of interest. For such systems, the coarse-grained entropy satisfies
\be
S \geq0,\quad\dot{S}\geq0,\quad\ddot{S}\leq0 \; . \label{Sconditions}
\ee
For example, consider a clump of particles, with some initial Gaussian density distribution, $\rho \sim \exp(-r^2/2)$, diffusing outwards with diffusion constant $D$. The diffusion equation implies that $\rho(r, t) = (2\pi(1+2Dt))^{-3/2} \exp \left ( - \frac{r^2}{2(1+2Dt)} \right )$.
It is then easy to check that the entropy, $S = - \int d V \rho \ln \rho$, obeys $\ddot{S} = - \frac{2}{3}\dot{S}^2$ at all times, so that (\ref{Sconditions}) holds. 

In fact, this is a very general property. As reviewed in the Appendix, it can be shown quite generally \cite{entropyproduction} that $\ddot{S} \leq 0$ for virtually all near-equilibrium systems approaching internal equilibrium. That is, finite, closed systems at late times inevitably obey (\ref{Sconditions}). By near-equilibrium, we mean systems that are characterized by $(\dot{S}/S)^2 \ll |\ddot{S}/S|$, which follows from $S \sim S_{\rm max}$ in this context. For systems that are not near equilibrium, $\ddot{S}$ can generically have either sign and hence (\ref{Sconditions}) may or may not hold; the diffusing gas is an example of a system in which (\ref{Sconditions}) does hold even though the system is never near equilibrium unless the gas is placed in a finite volume. But we emphasize that, as shown in the Appendix,  (\ref{Sconditions}) is guaranteed to hold for near-equilibrium systems.

Next, let us attempt to connect thermodynamics to local regions of spacetime. The motivation is as follows. The Bekenstein-Hawking entropy formula associates entropy to the area of black hole horizons. The formula is universal, applying to the horizons of all kinds of black holes in any number of dimensions. It even applies to de Sitter horizons. But most strikingly, the formula is also considered to hold (as an entropy density) for acceleration horizons. Since such horizons could be anywhere, this suggests that there might be a local entropy associated with the areas of patches of certain null surfaces. The idea of emergent gravity is to assume that this local entropy is similar to entropy in statistical-mechanical systems. That is, we assume that gravitational entropy arises as the coarse-grained entropy of some microscopic system of Planckian degrees of freedom associated with patches of certain null surfaces. What these degrees of freedom are is unknown and also largely irrelevant. It is not even clear whether these degrees of freedom live in spacetime or, because they have to account for an entropy that scales as an area, in some dual space in one lower dimension. We do know that for stationary horizons (including de Sitter and Rindler horizons), there is also an associated temperature. It therefore seems natural to assume that the underlying microscopic system is in fact a thermodynamic system. These two points are the basis for the idea that gravity might be described locally by some dual thermodynamic system. Despite little being known about the underlying system, the emergent gravity paradigm has met with great success due to Jacobson's remarkable result \cite{einsteineqnofstate} that Einstein's equations follow from what is essentially the first law of thermodynamics. Here, the only feature we will need to assume is that the underlying system either is already at, or is approaching, internal equilibrium via the second law of thermodynamics. 
Since the second law of thermodynamics is perhaps the most universal law in physics, this is not much of an assumption; we merely need to assume that the system is closed over the time-scales of interest. Moreover, since the idea is that the system is dual to an infinitesimal region of spacetime, the requirement that it be closed over infinitesimal times also seems natural.

Next, we would like to have a prescription for how to choose our null congruences. In Jacobson's paper, the thermodynamic system was taken to be instantaneously at equilibrium, and hence the corresponding null congruence was chosen to be a local Rindler horizon, with vanishing expansion and shear at the point of interest. Here we are interested in the second law, so we allow for non-equilibrium systems with increasing entropy. Correspondingly, we allow our congruences to have positive, or at least non-negative, local expansion. Our prescription then is very simple: we postulate that every non-contracting infinitesimal open patch of the integral curves of every null geodesic congruence is associated with a thermodynamic system obeying the second law; the
restriction to non-contracting patches enforces the second law of thermodynamics, which is the basic premise from which we will derive the null energy condition. Through a given spacetime point $p$ with a given future-directed null vector $v^\mu$ in the tangent space at $p$, there are infinitely many non-contracting geodesic congruences with tangent $v^\mu$ at $p$. We associate thermodynamic systems to {\em all} such infinitesimal patches. A particular class of expanding congruences consists of future light cones of earlier spacetime points. Among these, a special limiting case consists of the integral curves emanating from the future light cone of a point in the infinite past of $p$. Near $p$, the patch of such a stationary congruence is a local planar Rindler horizon, corresponding to an equilibrium system. Thus our prescription covers both equilibrium and non-equilibrium systems; it generalizes Jacobson's local Rindler horizons to patches whose local expansion can be not only zero, but also positive.

With this background, we identify the gravitational entropy of our infinitesimal patch with the coarse-grained entropy of a thermodynamic system. Then
\be
S = \frac{A}{4} \; .
\ee
It is implicit in this formula that classical physics is described by Einstein gravity minimally coupled to matter; for higher-curvature theories of gravity, or for non-minimally coupled gravity \cite{Chatterjee:2012}, the Bekenstein-Hawking entropy would have to be replaced by its appropriate generalization, such as the Wald entropy \cite{waldentropy}. Next, we identify the affine parameter of the null congruence with the time parameter in our thermodynamic system. Then
\be
\dot{S} = \frac{A}{4} \theta \; ,
\ee
and
\be
\ddot{S} = \frac{A}{4} \lf \theta^2 + \dot{\theta} \rt \; . \label{ddot}
\ee
Here we are assuming that $\theta$ is roughly constant over the surface; this is valid because the surface is infinitesimal. Notice that the near-equilibrium condition, $(\dot{S}/S)^2 \ll |\ddot{S}/S|$, translates to $\theta^2 \ll |\dot{\theta}|$.

Now because the congruence is null, its generators obey the optical Raychaudhuri equation:
\be
\dot{\theta}= - \frac{1}{2} \theta^2 - \sigma^2 + \omega^2 - R_{\mn} v^\mu v^\nu \; .
\ee
By hypersurface-orthogonality, $\omega^2 = 0$. The shear, $\sigma$, can always be chosen to vanish at a point. Choose an initial surface near or enclosing this point. In this region the shear will be small compared to $\theta$. Moreover, for small enough affine parameter $\lambda$ the shear will remain small compared to $\theta$. Then, for small times, $\sigma^2$ is negligible. 
We therefore drop the $\sigma$ and $\omega$ terms from Raychaudhuri's equation. Then we have
\begin{eqnarray}
R_{\mu \nu} v^\mu v^\nu & = & - ( \dot{\theta} + \theta^2 ) + \frac{1}{2} \theta^2 \nonumber \\
& = & - \frac{\ddot{S}}{S} + \frac{1}{2} \left (\frac{\dot{S}}{S} \right)^{\! \! 2} \label{RandS}
\end{eqnarray}
Now, for systems that are already at equilibrium, $\dot{S}$ and $\ddot{S}$ are both zero. Hence
\be
R_{\mu \nu} v^\mu v^\nu = 0 \; . \label{Rzero}
\ee
Next, consider systems approaching equilibrium. Then $\dot{S} > 0$. For systems that are far from equilibrium, $\ddot{S}$ can have either sign. Therefore, for expanding patches that correspond to far-from-equilibrium thermodynamic systems, the two terms on the right of (\ref{RandS}) could have different signs so that nothing can be inferred about the sign of $R_{\mu \nu} v^\mu v^\nu$ without knowing the precise values of $\dot{S}$ and $\ddot{S}$; no general statement can be made for such systems. However, for patches that correspond to near-equilibrium systems, we are guaranteed that  $\ddot{S} \leq 0$. The existence of such systems would guarantee that $R_{\mu \nu} v^\mu v^\nu \geq 0$. 

To complete the proof, we show existence of such congruences by construction. In the vicinity of the point $p$, $R_{\mu \nu} v^\mu v^\nu$ is a constant, namely $R_{\mu \nu}(p) v^\mu v^\nu$. Call this constant $C$. We will shortly determine the sign of $C$ from thermodynamics. Solving the Raychaudhuri equation for a shear-free congruence, we find
\be
\theta = \sqrt{2C} \tan \left (-\sqrt{\frac{C}{2}} \lambda + b \right ) \; ,
\ee
where $b$ is a constant of integration; different choices of $b$ correspond to different congruences. Choosing $b = 0$, we see that $\theta$ vanishes for $\lambda = 0$. Suppose we consider some open patch for very small $\lambda$ (but not including the point $\lambda = 0$, where the sign of $\theta$ changes). Then
\be
\theta \approx - C \lambda \comma \dot{\theta} \approx -C \; .
\ee
If $\theta = \dot{\theta} = 0$ then $C = 0$; stationary (equilibrium) congruences require (\ref{Rzero}). Otherwise, since $\lambda$ is chosen to be small, we see that $\theta^2 \ll |\dot{\theta}|$. This translates to $(\dot{S}/S)^2 \ll |\ddot{S}/S|$, which means that the system is indeed near equilibrium. We have thus shown, by explicit solution of the Raychaudhuri equation, that congruences corresponding to stationary (equilibrium) or near-equilibrium systems exist everywhere.

But if the system is near equilibrium, then we know from statistical mechanics that $\ddot{S} < 0$. By (\ref{ddot}), this in turn means $\dot{\theta} < 0$, so that $C > 0$, which is to say
\be
R_{\mu \nu} v^\mu v^\nu > 0 \; .
\ee
Therefore, for both equilibrium and non-equilibrium thermodynamic systems, we find $R_{\mu \nu} v^\mu v^\nu \geq 0$. This is precisely the geometric form of the null energy condition, (\ref{RNEC}). Since $v^\mu$ is any arbitrary future-directed null vector, this establishes the null energy condition.

\emph{Discussion.} \textendash{} The null energy condition was initially proposed as a plausible but ad hoc requirement on matter. This condition, which does not seem to follow from any first principles, has sweeping consequences when matter is coupled to gravity. Here we have taken a different view: we regard the null energy condition not as an ad hoc characteristic of matter, but as a fundamental property of gravity. Moreover, we have shown that this property, in the form of the Ricci convergence condition, follows directly from an assumption that some underlying conventional non-gravitational microphysics accounts for the Bekenstein-Hawking entropy and obeys the second law of thermodynamics. It is remarkable that the point-wise classical null energy condition, which in its matter form has so far been impossible to derive from quantum field theory, follows in its geometric form so readily from the thermodynamics of emergent gravity. It is a satisfying result because the universality of the null energy condition -- which is supposed to hold for all physical spacetimes -- is traced to another universal condition, namely the second law of thermodynamics. 

In this work, our underlying premise has been that all non-contracting infinitesimal open patches of the integral curves of null geodesic congruences can be associated with thermodynamic systems. How then, should we interpret geodesic congruences that are locally contracting? One can imagine several alternatives. First, it may well be that the existence of congruences with $\theta < 0$ (or in which $\theta$ changes sign) merely indicates that our premise is wrong. This is certainly a logical possibility. But the same critique could be applied to Jacobson's original paper, which restricts discussion to patches of null congruences with vanishing $\theta$ (``local Rindler horizons"), an even more restrictive set of congruences than the one we consider. In both cases, however, accepting the premise leads to a non-trivial result (Einstein's equations, null energy condition). Perhaps one could regard this as evidence for the assumption. Second, it may be that the correct way to associate thermodynamics with geometry is to start from the microscopic system. In this case, not every geometric surface or congruence need correspond to something that has a meaningful microscopic interpretation. In this approach, if we start with microscopic thermodynamic systems that obey the second law, we should necessarily consider only null congruences with $\theta \geq 0$, and we need not inquire about the interpretation of other congruences. Third, it may be that all congruences, even those with $\theta < 0$, do in fact correspond to thermodynamic systems. For suppose we have a contracting patch. We could simply identify thermodynamic time with negative affine parameter, $\lambda$. Then $\theta < 0$ would still correspond to $\dot{S} > 0$. The Raychaudhuri equation is invariant under $\lambda \leftrightarrow - \lambda$, and so we would still obtain the null energy condition as a consequence of thermodynamics; in this way, patches in which $\theta < 0$ can be accommodated as well. That leaves only patches for which $\theta$ changes sign. But these are rare events of measure zero; one can speculate that these may correspond to rare violations of the second law. 

Finally, it is striking that there are two distinct derivations of the null energy condition, from worldsheet string theory \cite{NECderivation} as well as from thermodynamics in the emergent gravity paradigm, and it would surely be illuminating to understand why two derivations exist \cite{tworoads}. Another interesting question is whether this calculation can be extended to higher-derivative gravity. Indeed it has been non-trivial to derive the generalized Einstein's equation from thermodynamics \cite{beyond,guedens}. In higher-derivative gravity, (\ref{TNEC}) and (\ref{RNEC}) are no longer equivalent so it is not clear what the correct condition is \cite{Chatterjee:2012,necviolating}. Perhaps the approach here will point the way.

\bigskip
\noindent
{\bf Acknowledgments}
\noindent
We thank Ted Jacobson, Cynthia Keeler, Erik Verlinde, and Frank Wilczek for helpful discussions. MP is supported in part by John Templeton Foundation grant 60253.

\bigskip
\noindent
\emph{Appendix: $\ddot{S}\leq0$.} \textendash{} Here we reproduce a proof of Falkovich and Fouxon \cite{entropyproduction} showing that typical near-equilibrium thermodynamic systems relaxing to equilibrium must have $\ddot{S} \leq 0$. 
Consider a phase space density $\rho$ associated with a reduced description of the system (due to coarse-graining). Suppose the system is close to thermodynamic equilibrium. Then the phase space density is near  the value $\rho_{0}$ that maximizes the entropy:
\be
\rho=\rho_{0}+\delta \rho \; .
\ee
Then
\begin{eqnarray}
S(\rho_{0}+\delta \rho)&=&-\int (\rho_{0}+\delta \rho)\ln(\rho_{0}+\delta \rho) \nonumber \\
&\approx& S_{\rm max}-\int \left ( \rho_{0}^{-1}\frac{(\delta \rho)^{2}}{2}\right ) \; ,
\end{eqnarray}
where $S_{\rm max}=-\int  \rho_{0}\ln \rho_{0} $
and we have used the fact that $\delta S|_{\rho_{0}}=0$. Near equilibrium, the time-derivative of
the density fluctuation satisfies a linear Onsager relation:
\be
\delta\dot{\rho}=\hat{L}\delta \rho \; ,
\ee
where the Onsager $\hat{L}$ matrix is taken to be symmetric. As Onsager showed \cite{onsager}, the symmetry of $\hat{L}$ follows from the principle of microscopic reversibility, so long as the macroscopic thermodynamic state variables are themselves time-invariant; this is the case for all but a few ``exceptional" systems of interest (usually involving magnetic fields). It seems quite likely that the thermodynamics of the microscopic theory of gravity satisfies these time-invariance properties; here we assume that this is the case. ($\hat{L}$ is presumably also invariant under time-translations.) When $\hat{L}$ is symmetric, we can expand $\delta \rho$ into orthonormal eigenfunctions of $\hat{L}$:
\be
\delta \rho=\sum_{k}\sqrt{\rho_{0}}a_{k}\psi_{k} \; ,
\ee
where $\hat{L}\psi_{k}=\lambda_{k}\psi_{k}$. 
Now
\be
\dot{S}=-\int \rho_{0}^{-1}\delta \rho (\hat{L}\delta \rho)  \; .
\ee
Then the second law implies
\be
-\sum_{j,k}\int \left(a_{j}a_{k}\lambda_{k}\psi_{j}\psi_{k}\right) \geq0\Rightarrow\lambda_{k}\leq 0 \; , \label{lambdak}
\ee
for all $k$. That is, the second law indicates that the eigenvalues of the operator $\hat{L}$ are real (and non-positive). 
Now consider the second derivative:
\begin{eqnarray}
\ddot{S} & =&-\int \rho_{0}^{-1}\left[\delta\dot{\rho}(\hat{L}\delta \rho)+\delta \rho (\hat{L}\delta \dot{\rho})\right] \nonumber\\
&=&-\int  \rho_{0}^{-1}\left[\left(\hat{L}\delta \rho \right)^{\! 2}+\delta \rho \left(\hat{L}^{2}\delta \rho \right)\right] \; .
\end{eqnarray}
Inserting the eigenfunction expansion, we find
\be
\ddot{S} 
=-2\sum_{k}a_{k}^{2}\lambda_{k}^{2} \; ,
\ee
so that
\be
\ddot{S} \leq0 \; .
\ee
Note from (\ref{lambdak}) that if $\dot{S} = 0$ then $\ddot{S} = 0$ while if $\dot{S} > 0$ then $\ddot{S} < 0$.



\begin{thebibliography}{99}
\bibitem{derivatives}
For a closed system, the same argument would naively also lead to an infinite number of conditions on higher time-derivatives of entropy. However, specification of all time derivatives would imply knowing the exact form of the entropy for all time. But the system can be regarded as closed only over short time-scales so extending the argument to all time derivatives is not justifiable.
\end{thebibliography}
\end{document}